\documentclass[a4paper,11pt]{article}
\usepackage{amsfonts}
\usepackage{amsmath}
\usepackage{amssymb}
\usepackage{graphicx}
\usepackage{array}
\usepackage{booktabs}
\usepackage{multirow}
\usepackage{makecell}
\usepackage{float}
\pdfoutput=1 

\usepackage{mymacros}
\usepackage{chngcntr}
\usepackage{verbatim}
\usepackage{amsthm}
\usepackage{graphics}
\usepackage{xcolor}
\usepackage{mathrsfs}
\usepackage{bbold}
\usepackage{cases}
\usepackage{epsfig}
\usepackage{epstopdf}
\usepackage{hyperref}
\usepackage{amsfonts}
\usepackage{hyperref}
\usepackage{jheppub} 

\usepackage[T1]{fontenc} 

\hypersetup{
	colorlinks=true,
	linkcolor=blue,
	filecolor=blue,
	urlcolor=blue,
	citecolor=blue,
}

\title{\boldmath Lyapunov exponents as probes for a phase transition of a  Kerr-AdS black hole}

\author{Deyou Chen,}
\author{Chuang Yang,}
\author{Yongtao Liu}

\emailAdd{deyouchen@hotmail.com}
\emailAdd{chuangyangyc@hotmail.com}
\emailAdd{lytao@foxmail.com}

\affiliation{School of Science, Xihua University, Chengdu 610039, China}

\abstract{In this letter, we study proper time Lyapunov exponents and coordinate time Lyapunov exponents of chaos for both massless and massive particles orbiting a four-dimensional Kerr-AdS black hole, and explore their relationships with the phase transition of this black hole. The result reveals that both types of Lyapunov exponents can reflect the occurrence of the phase transition.	 Specifically, when compared to the proper time and coordinate time Lyapunov exponents of massive particles in chaotic states, the exponents corresponding to massless particles demonstrate a more robust capability in describing the phase transition. Furthermore, we conduct a study on critical exponents associated with these Lyapunov exponents in this black hole, identifying a critical exponent value of 1/2.}


\begin{document} 
	\maketitle
	\flushbottom
	
\section{Introduction}

The motions of particles in the vicinity of black holes(BHs) has garnered significant attention, as these motions convey important information about their background spacetimes. Specifically, particles possess an innermost stable circular orbit around a BH, and once they cross this critical threshold, they are inevitably swallowed by the BH \cite{PQR,LLL,ZWSYL,ZL}. The radius of this orbit is contingent upon the BH's mass and rotational velocity, thereby serving as a reflection of its spacetime geometry. Furthermore, the study of these orbits provides valuable insights into accretion disks and associated radiation spectra \cite{PT}. 

The unique physical properties and spacetime structures of BHs can induce instability in the geodesics of particles orbiting them. Compared to stable geodesics, these unstable geodesics possess greater observational value. BHs' shadows are cast by combined effects of their event horizons' capture of photons and strong gravitational lensing effect \cite{EHT1,EHT2,EHT3,EHT4}. By observing these shadows, one can deduce fundamental attributes such as the mass and rotation of the BHs, as well as the physical characteristics of their surrounding environments, thereby constraining related physical parameters. Furthermore, observations of BH shadows hold potential for shedding light on BH mergers. When disturbed by external fields, BHs evolve and propagate outward in the form of quasinormal modes(QNMs). The vibration frequencies and decay times of these QNMs are governed by the BHs' characteristic parameters, including their masses, charges and angular momenta. Null geodesics serve as important tools for acquiring these QNMs \cite{RAK}. Specifically, the angular velocities at the unstable null geodesics orbits determine the real parts of the QNMs, whereas their imaginary parts are associated with the unstable time scales of the orbits. Notably, the dominant modes can be observed in the gravitational wave signals emitted by BHs \cite{BPA1,BPA2,BPA3}. Optical appearances of stars undergoing gravitational collapse largely depend on the circular unstable null geodesics, which also explain the exponential fade-out of the luminosity of the collapsing stars \cite{AT}. For some unstable geodesics, even slight disturbances can trigger chaotic motion of particles \cite{KT11,KT12,KT13}. This chaotic signature would leave an imprint on gravitational waves emitted from BHs \cite{KT1,KT2,KT3,KT4,KT5,KT6}. Therefore, the study of unstable geodesics around BHs not only provides insights into the fundamental properties of BHs but also offers potential observational signatures that can be detected through gravitational wave observations and other astronomical phenomena.

On the other hand, the exploration of phase transitions in BHs originated with the revelation of the Hawking-Page phase transition \cite{SWHP}. Following the interpretation of the cosmological constant in anti-de Sitter(AdS) spacetimes as pressure, a series of works unveiled striking similarities between BHs and fluids in the context of phase transitions \cite{AA1,AA2,AA3,AA4,AA5,AA6,AA7,AA8,AA9,AA10,BB1,BB2,BB3,BB4,BB5,BB6,BB7,BB8,BB9,BB10,CC1,CC2,CC3,CC4,CC5,CC6}. During these phase transitions, the BHs parameters change, which can directly impact on the physical environment surrounding them. This alteration in physical environment may trigger the occurrence of chaos among the particles  \cite{DDS}. Furthermore, the chaos of particles around BHs may subtly alter the states of the systems, which can accumulate and eventually trigger phase transitions. Consequently, there exists a close connection between the chaos and phase transitions, where the chaotic behaviors of particles can reflect the phase transitions in the systems. In light of this connection, the Lyapunov exponent(LE) in terms of coordinate time associated with the chaotic behavior of the particles and ring strings around the Reissner-Nordstr\"om-AdS BH was researched in \cite{GLMW}. The study revealed that the exponent exhibits multiple values when a phase transition occurs. The discontinuous change in the exponent was considered as an order parameter, and a relationship between this exponent and the critical temperature was established, yielding a critical exponent of 1/2. Subsequently, this work was extended to other spherically symmetric spacetimes \cite{YTMH,LTW,KPA,DLMG,SDDM,GAP}. These studies provide novel insights into the understanding of phase transitions. However, these studies have primarily concentrated on the relationship between the LEs in terms of coordinate time and phase transitions of spherically symmetric BHs. It is equally significant to study the connection between LEs in terms of proper time and phase transitions of axisymmetric BHs. This endeavor promises to further deepen our comprehension of the interplay between chaos and phase transitions in BH systems.

In this letter, we study proper time Lyapunov exponents(PTLEs) and coordinate time Lyapunov exponents(CTLEs) of chaotic motion exhibited by massless and massive particles around a  four-dimensional Kerr-AdS BH in canonical ensemble, and explore their relationships with the phase transition of this BH. In the calculation, we fix these particles' angular momenta and calculate the critical value of the angular momentum of the BH at the phase transition point. Building upon this foundation, we further examine these relationships under scenarios where the angular momentum falls below its critical threshold. Furthermore, we compute the critical exponents which are related to both types of LEs.

The remainder of this letter is organized as follows. In the subsequent section, we review the acquisition of the PTLE and CTLE. In Section \ref{sec3}, we study the phase transition of the four-dimensional Kerr-AdS BH, exploring its relationship with the PTLEs and CTLEs of both massless and massive particles in their chaotic motions. Furthermore, we calculate the critical exponents associated with both the CTLEs and PTLEs. The final section is devoted to our conclusions.

\section{Review of LEs}\label{sec2}

LEs serve as significant indicators for quantifying chaotic characteristics, as they characterize the average exponential rates at which adjacent orbits in the phase space of classical systems either converge or diverge. A positive LE signifies divergence between two neighboring geodesics, thereby indicating that the system is highly sensitive to initial conditions, potentially leading to chaotic behavior. Conversely, a negative exponent implies a tendency towards stability within the system. When the exponent is zero, the system exhibits periodic motion. In this section, we first review the acquisition of both the PTLE and CTLE \cite{CMBWZ,PP1,PP2,PP3,LG1,LG2,GCYW}.

For a particle in equilibrium outside a BH, its equation of motion can be schematically represented as follows
 
\begin{eqnarray}
	\frac{dX_i}{dt} = F_i(X^j),
	\label{eq2.1.1}
\end{eqnarray}

\noindent where $X_i$ are coordinates and $F_i(X^j)$ are functions to be determined. We linearise the above equations around a certain orbit and get

\begin{eqnarray}
	\frac{d\delta X_i(t)}{dt} =K_{ij}(t)\delta X_j(t),
	\label{eq2.1.2}
\end{eqnarray}

\noindent where $K_{ij}(t) $ is a Jacobian matrix defined by

\begin{eqnarray}
	K_{ij}(t) = \left.\frac{\partial F_i}{\partial X_j}\right|_{X_i(t)}.
	\label{eq2.1.3}
\end{eqnarray}

\noindent When the particle moves in an circular orbit with radius $r_0$ on the equator of the black hole, we define the classical phase space variables $X_i(t) = (p_r, r)$.  Evidently, such an orbit can exist in spherically symmetric spacetimes or in the equatorial planes of axisymmetric spacetimes \cite{CMBWZ}. Utilizing the canonical momenta, given by $p_q = \frac{\partial \mathcal{L}}{\partial \dot{q}}$, and the Euler-Lagrange equations of motion 

\begin{eqnarray}
\frac{dp_q}{d\tau} = \frac{\partial \mathcal{L}}{\partial q},
\label{eq2.1.4}
\end{eqnarray}

\noindent we get $\frac{dp_r}{d\tau} = \frac{\partial \mathcal{L}}{\partial r}$ and $\frac{dr}{d\tau} = \frac{p_{r}}{g_{rr}}$. Consequently, the matrix is obtained as
 
\begin{eqnarray}
K_{ij}=\begin{pmatrix}
0 & \frac{d}{dr}\left(\frac{\partial \mathcal{L}}{\partial r}\right)\\
\frac{1}{g_{rr}} & 0
\end{pmatrix}.
\label{eq2.1.5}
\end{eqnarray}

\noindent The exponent is determined by the eigenvalues of the Jacobian matrix evaluated at the equilibrium point, which are obtained as follows

\begin{eqnarray}
	\lambda^2_p = \frac{1}{g_{rr}}\frac{d}{dr}\left(\frac{\partial \mathcal{L}}{\partial r}\right),
	\label{eq2.1.6}
\end{eqnarray}

\noindent To further derive its expression, we utilize Lagrange's equation of geodesic motion

\begin{eqnarray}
\frac{d}{d\tau}\left(\frac{\partial\mathcal{L}}{\partial \dot{r} }\right) - \frac{\partial \mathcal{L}}{\partial r}=0,
\label{eq2.1.7}
\end{eqnarray}

\noindent and the formula

\begin{eqnarray}
\frac{d}{d\tau}\left(\frac{\partial\mathcal{L}}{\partial \dot{r} }\right) =\frac{d}{d{\tau}}(g_{rr}\dot{r})= \frac{1}{2g_{rr}}\frac{d}{dr}(g_{rr}^2 \dot{r}^2).
\label{eq2.1.8}
\end{eqnarray}

\noindent Thus the expression of $\frac{\partial\mathcal{L}}{\partial r}$ is gotten 

\begin{eqnarray}
\frac{\partial\mathcal{L}}{\partial r} = \frac{1}{2g_{rr}}\frac{d}{dr}(g_{rr}^2 \dot{r}^2).
\label{eq2.1.9}
\end{eqnarray}

\noindent After defining an effective potential as $V_r=\dot{r}^2$, we substitute this formula into Eq. (\ref{eq2.1.6}). Taking into account an unstable equilibrium orbit, we find that $V_r=V^{\prime}_r=0$, where the prime denotes the derivative with respect to $r$. Consequently, the exponent is obtained as outlined below

\begin{eqnarray}
\lambda_p = \sqrt{\frac{ V^{\prime\prime}_r}{2}},
\label{eq2.1.10}
\end{eqnarray}

\noindent which is defined as the PTLE. On the other hand, from the Euler-Lagrange equations of motion, we have $\frac{dp_r}{dt} = \frac{d\tau}{dt}\frac{\partial \mathcal{L}}{\partial r}$. Thus the exponent in term of coordinate time is expressed as \cite{LG1}

\begin{eqnarray}
\lambda_c = \sqrt{\frac{ V^{\prime\prime}_r}{2\dot{t}^2}},
\label{eq2.1.11}
\end{eqnarray}

\noindent which is defined as the CTLE. These two exponents reflect the stability of the motion of equatorial particles in the spherical spacetimes and in the equatorial planes of the axisymmetric spacetimes. Their relationship is expressed by $\lambda_p=|\dot{t}|\lambda_c$. Since $|\dot{t}|>0$, their properties are closely related in the spherically symmetric spacetimes \cite{LG1}.

\section{LEs and phase transition of four-dimensional Kerr-AdS BH}\label{sec3}

\subsection{Thermodynamics of four-dimensional Kerr-AdS BH}\label{sec3.1}

The Kerr-AdS BH is a vacuum solution of Einstein field equations characterized by a negative cosmological constant. It describes a rotational AdS spacetime and its metric is given by 

\begin{eqnarray}
ds^2 =-\frac{\Delta}{\rho^2}\left(dt-\frac{a\sin^2\theta}{\Xi} d\varphi\right)^2 +\frac{\rho^2}{\Delta}dr^2 + \frac{\rho^2}{\Sigma} d\theta^2  +\frac{\Sigma\sin^2\theta}{\rho^2}\left[adt -\frac{(r^2+a^2)}{\Xi}d\varphi\right]^2,
\label{eq2.2.1}
\end{eqnarray}

\noindent with the metric functions

\begin{eqnarray}
\Delta &=& \left(r^2+a^2\right)\left(1+\frac{r^2}{l^2}\right) -2mr, \quad \Xi= 1-\frac{a^2}{l^2}, \nonumber\\
\rho^2 &=& r^2 + a^2\cos^2\theta,\quad \Sigma= 1-\frac{a^2}{l^2}\cos^2\theta.
\label{eq2.2.2}
\end{eqnarray}

\noindent where $m$, $a$ and $l$ are the mass parameter, rotational parameter and AdS radius, respectively. $l$ is related to the cosmological constant as $\Lambda =-3l^2$. The metric is singular when $a^2 = l^2$. For both $a^2 < l^2$ and $a^2 > l^2$, the metric can exist, and the metric for the latter can be mapped to that for the first by a scaling transformation \cite{CLP}. The ADM mass and the angular momentum are 

\begin{eqnarray}
M=\frac{m}{\Xi^2}, \quad J=\frac{ma}{\Xi^2}.
\label{eq2.2.3}
\end{eqnarray}

\noindent The entropy and Hawking temperature are

\begin{eqnarray}
S &=& \frac{\pi(r_+^2 + a^2)}{\Xi},\\
T &=& \frac{3r_+^4+r_+^2(a^2+l^2)-a^2l^2}{4\pi l^2 r_+(r_+^2+a^2)},
\label{eq2.2.4}
\end{eqnarray}

\noindent where $r_+$ is the event horizon determined by the largest positive root of $\Delta=0$. At the horizon, the angular velocity is  

\begin{eqnarray}
\Omega_+ = \frac{a\Xi}{r_+^2+a^2}.
\label{eq2.2.5}
\end{eqnarray}

\noindent In the extended phase spaces, the cosmological constant is treated as a variable to pressure, while the mass is regarded as enthalpy. The phase transition and thermodynamics of this BH has been discussed in \cite{AKMS1,AKMS2,YZWL1,YZWL2,YZWL3}. In these studies, the pressure is defined as $P=\frac{3}{8\pi l^2}$, and the conjugate thermodynamic volume is $V = \frac{4\pi(r_+^2 + a^2)}{3\Xi}\left[\frac{(r_+^2+l^2)a^2}{2r_+^2\Xi l^2}+1\right]$. The Gibbs free energy for the BH is

\begin{eqnarray}
F = M-TS = \frac{(r_+^2 + a^2)(r_+^2+l^2)}{2\Xi^2 l^2r_+}- \frac{3r_+^4+r_+^2(a^2+l^2)-a^2l^2}{4\Xi l^2 r_+}.
\label{eq2.2.7}
\end{eqnarray}

\begin{figure}[h]
	\centering
	\includegraphics[width=10cm,height=7cm]{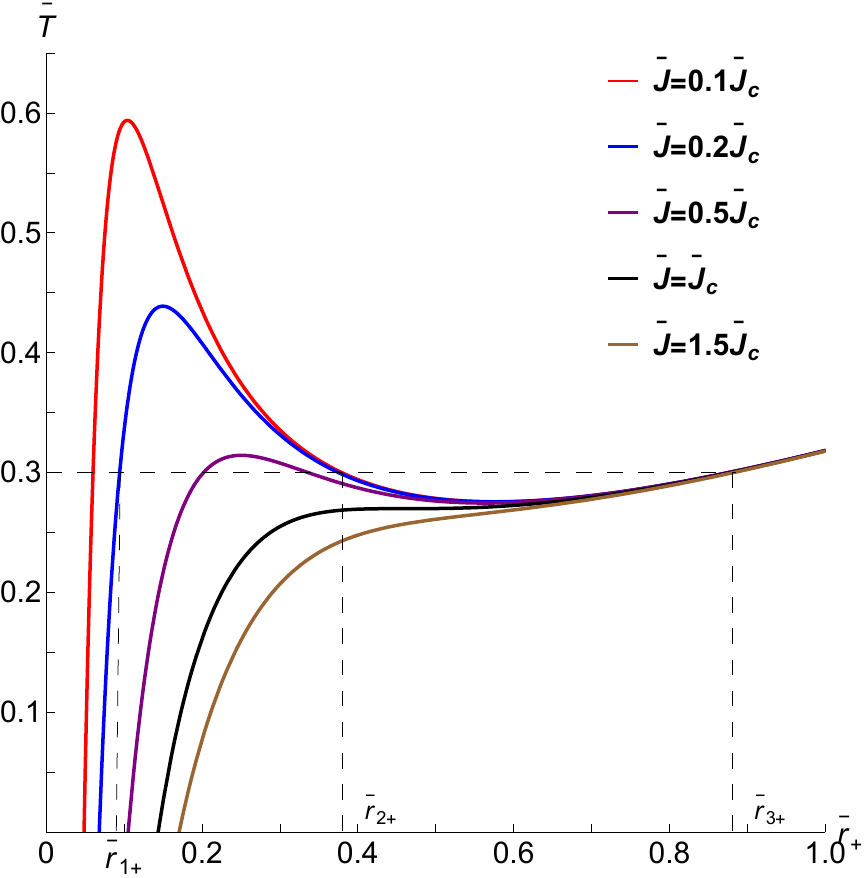}
	\caption{The temperature of the four-dimensional Kerr-AdS BH varies as a function of the horizon radius.}
	\label{4f1}
\end{figure}

Through dimensional analysis, we observe that certain physical quantities scale as powers of  $l$ and they can be written as follows 

\begin{eqnarray}
r =\bar{r}l, \quad r_+ =\bar{r}_+l,  \quad a= \bar{a}l, \quad m= \bar{m}l, \quad J= \bar{J}l^2, \quad T = \bar{T}l^{-1}, \quad F= \bar{F}l.
\label{eq2.2.8}
\end{eqnarray}

\noindent where $\bar{r}$, $\bar{r}_+$, $\bar{a}$, $\bar{m}$, $\bar{J}$, $\bar{T}$ and $\bar{F}$  are all dimensionless quantities. Eq. (\ref{eq2.2.4}) demonstrates that the temperature $T$ is a function of $r_+$.  By utilizing this equation in conjunction with Eqs. (\ref{eq2.2.3}) and (\ref{eq2.2.8}),  we can derive a function $\bar{r}_+(\bar{T})$. When a specific value of $\bar{T}$ corresponds to multiple values of $\bar{r}_+$, it signifies that the BH possesses multiple phases. Conversely, when this is not the case, the BH displays only a single phase. This relationship is depicted in Figure \ref{4f1},  where $\bar{J}_c=0.023933$ represents the value at a critical point and is determined by 

\begin{eqnarray}
\left(\frac{\partial \bar{T}}{\partial \bar{r}_+}\right)_{\bar{J}}=\left(\frac{\partial^2 \bar{T}}{\partial \bar{r}_+^2}\right)_{\bar{J}}=0.
\label{eq2.2.9}
\end{eqnarray}

\noindent The other critical values are $\bar{r}_{+c} = 0.458819$ and $\bar{T}_c=  0.269866$. In Figure \ref{4f1}, it is observed that there is no inflection point, and the temperature increases monotonically for $\bar{J} =1.5\bar{J}_c$. This indicates that each phase is uniquely associated with a single temperature value. Conversely, when $\bar{J}<\bar{J}_c$, inflection points appear, suggesting that the BH exhibits multiple phases for a given value of $\bar{T}$. Specifically, for $\bar{J}=0.2\bar{J}_c$ and $\bar{T} =0.3$, there exist three roots: $\bar{r}_{1+}$, $\bar{r}_{2+}$ and $\bar{r}_{3+}$, which correspond to three phases of small, intermediate and large BHs, respectively.

\begin{figure}[h]
	\centering
	\begin{minipage}[t]{0.48\textwidth}
		\centering
		\includegraphics[width=7cm,height=5cm]{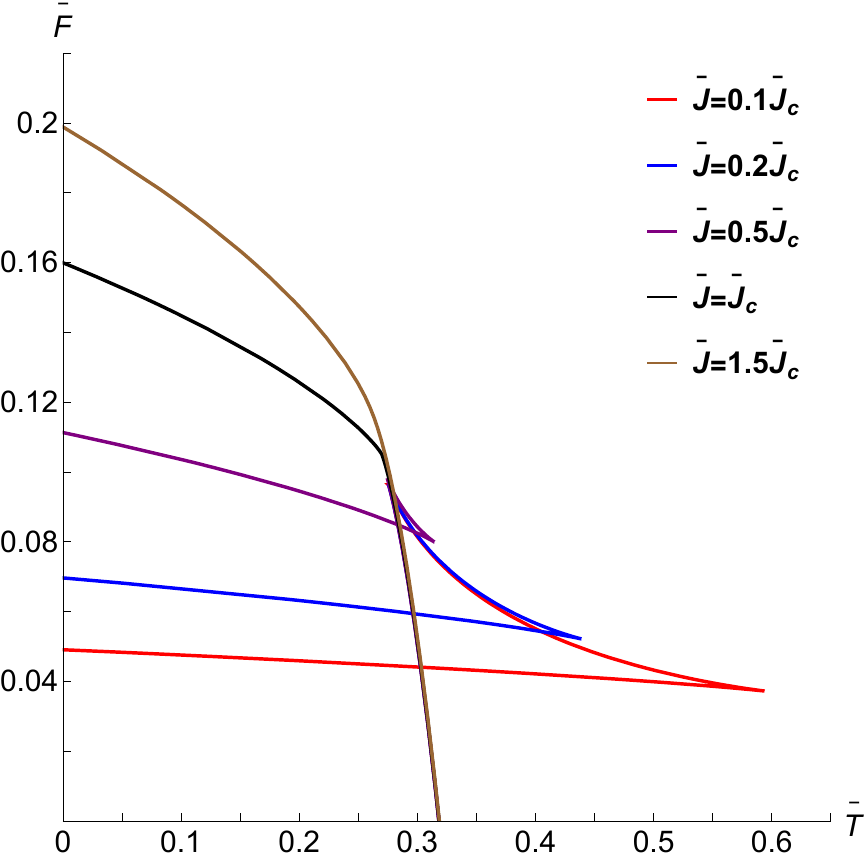}
		\subcaption{}
		\label{4-2-a}
	\end{minipage}
	\begin{minipage}[t]{0.48\textwidth}
		\centering
		\includegraphics[width=7cm,height=5cm]{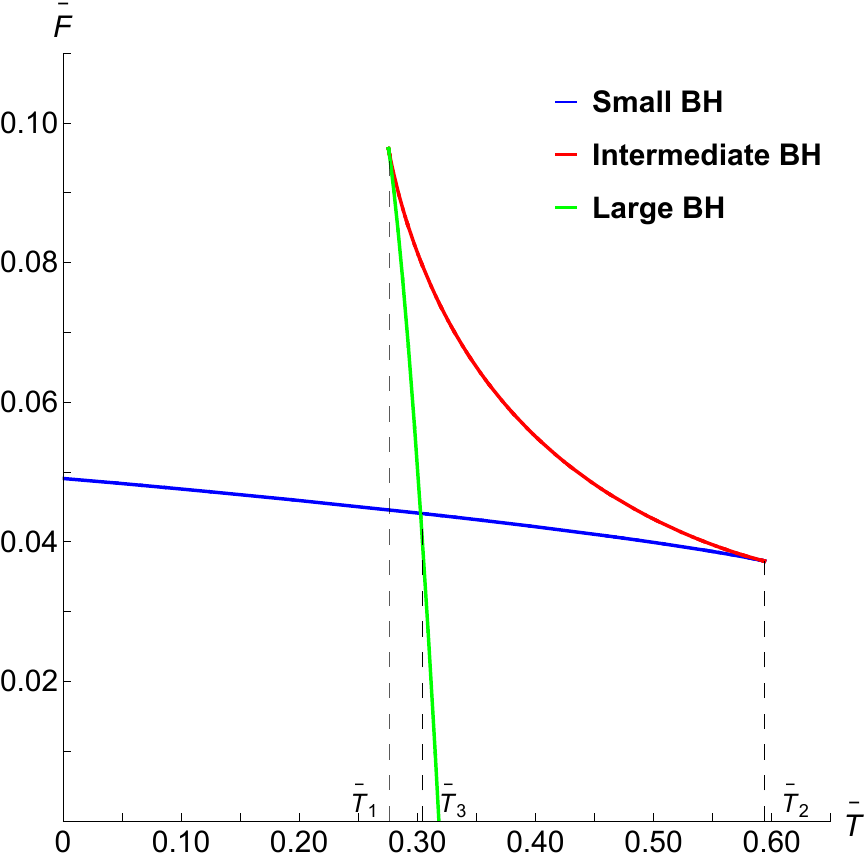}
		\subcaption{}
		\label{4-2-b}
	\end{minipage}
	\caption{The free energy of the four-dimensional Kerr-AdS BH varies with the temperature. The scenario where $\bar{J} =0.1\bar{J}_c$ is illustrated in Figure \ref{4-2-b}.}
	\label{4f2}
\end{figure}

To further elucidate the phase transition, we present the curves depicting the free energy as a function of the temperature in Figure \ref{4f2}. In Figure \ref{4-2-a}, the absence of a swallowtail-like structure for $\bar{J} =1.5\bar{J}_c$ indicates the absence of a phase transition. Conversely, when $\bar{J} <\bar{J}_c$, swallowtail-like structures emerge, typically signaling the onset of a phase transition between large and small BHs. Figure \ref{4-2-b} specifically illustrates the change in free energy at $\bar{J} =0.1\bar{J}_c$. Here, the free energy monotonically decreases for $\bar{T} <\bar{T}_1$ and $\bar{T} > \bar{T}_3$. A small BH phase is observed for $\bar{T} <\bar{T}_1$, while a large BH phase appears for $\bar{T} >\bar{T}_3$. In both cases, the BHs are stable. Within the range $\bar{T}_1<\bar{T} <\bar{T}_3$, the free energy becomes multivalued, with small, intermediate and large BHs coexisting. These phases transform into each other within this temperature range. The first-order phase transition between a small and large BH occurs at  $\bar{T}_3$.

\subsection{LEs and phase transition}\label{sec3.2}

The motions of particles outside BHs can reflect certain characteristics of the background spacetimes in which the particles reside. In this section, we study the thermodynamic phase transition of the four-dimensional Kerr-AdS BH by examining the PTLE and CTLE associated with the chaotic motion of both massless and massive particles.

When a neutral particle moves in the equatorial plane of the BH, its Lagrangian is given by

\begin{eqnarray}
\mathcal{L} = \frac{1}{2}g_{\mu\nu} \dot{x}^{\mu}\dot{x}^{\nu}= \frac{1}{2}\left(\bar{g}_{tt}\dot{t}^2+\bar{g}_{rr}\dot{r}^2 +\bar{g}_{\varphi\varphi}\dot{\varphi}^2+2\bar{g}_{t\varphi}\dot{t}\dot{\varphi}\right).
\label{eq2.2.10}
\end{eqnarray}

\noindent Here we use $\bar{g}_{\mu\nu}$, where $\mu ,\nu= t,r,\varphi$, to represent the components of the Kerr-AdS metric in the equatorial plane ($\theta = \frac{\pi}{2}$). Dots are employed to denote derivatives with respect to the proper time. The specific expressions for $\bar{g}_{\mu\nu}$ are 

\begin{eqnarray}
\bar{g}_{tt} = -\frac{\Delta - a^2}{r^2},\quad
\bar{g}_{rr} =  \frac{r^2}{\Delta},\quad
\bar{g}_{\varphi\varphi}  = \frac{(r^2+a^2)^2 -\Delta a^2}{r^2\Xi^2},\quad
\bar{g}_{t\varphi} = -\frac{(r^2+a^2-\Delta)a}{r^2\Xi}.
\label{eq2.2.12}
\end{eqnarray}

\noindent Using the definition of generalized momenta $p_{\mu}=\frac{\partial\mathcal{L}}{\partial\dot{x}^{\mu}} = g_{\mu\nu}\dot{x}^{\nu}$, we get

\begin{eqnarray}
p_t &=& \bar{g}_{tt}\dot{t}+ \bar{g}_{t\varphi}\dot{\varphi}=-E,\\
p_r &=& \bar{g}_{rr}\dot{r}, \\
p_{\varphi} &=& \bar{g}_{\varphi\varphi}\dot{\varphi}+ \bar{g}_{t\varphi}\dot{t} =L,
\label{eq2.2.14}
\end{eqnarray}

\noindent where $E$ and $L$ are the energy and angular momentum of the particle, respectively. From the above equations, it is straightforward to derive the equations governing $t$-motion and $\varphi$-motion,

\begin{eqnarray}
\dot{t} &=& \frac{E \bar{g}_{\varphi\varphi}+L\bar{g}_{t\varphi}}{g_{t\varphi}^2-\bar{g}_{tt}\bar{g}_{\varphi\varphi}}, \\
\dot{\varphi} &=& \frac{E\bar{g}_{t\varphi}+L\bar{g}_{tt}}{\bar{g}_{tt}g_{\varphi\varphi}-\bar{g}_{t\varphi}^2}.
\label{eq2.2.16}
\end{eqnarray}

\noindent The Hamiltonian of the system is 

\begin{eqnarray}
2\mathcal{H} = 2(p_{\mu}\dot{x}^{\mu}-\mathcal{L}) = p_{\mu}\dot{x}^{\mu}= \bar{g}_{tt}\dot{t}^2+\bar{g}_{rr}\dot{r}^2 +\bar{g}_{\varphi\varphi}\dot{\varphi}^2+2\bar{g}_{t\varphi}\dot{t}\dot{\varphi}=\delta,
\label{eq2.2.17}
\end{eqnarray}

\noindent where $\delta =0$ and $-1$ correspond to null and time-like geodesics, respectively. Solving Eq. (\ref{eq2.2.17}), we get $r$-motion

\begin{eqnarray}
\dot{r}^2 = \frac{\delta}{\bar{g}_{rr}}-\frac{E^2 \bar{g}_{\varphi\varphi} +L^2 \bar{g}_{tt} +2EL\bar{g}_{t\varphi} }{\bar{g}_{rr}(\bar{g}_{tt}\bar{g}_{\varphi\varphi}-\bar{g}_{t\varphi}^2)}.
\label{eq2.2.18}
\end{eqnarray}

\subsubsection{Null geodesic's case}\label{sec3.2.1}

The motion of massless particles is governed by the null geodesic equations. By employing the definition of the effective potential, denoted as $V_r = \dot{r}^2$, we derive an expression for the effective potential of a massless particle within the context of Kerr-AdS spacetime,

\begin{eqnarray}
V_{r}= \frac{(aE-L\Xi)^2\Delta-\left[E(r^2+a^2)-aL\Xi_a)\right]^2}{r^4}.
\label{eq2.3.1}
\end{eqnarray}

\begin{figure}[h]
	\centering
	\begin{minipage}[t]{0.48\textwidth}
		\centering
		\includegraphics[width=7cm,height=5cm]{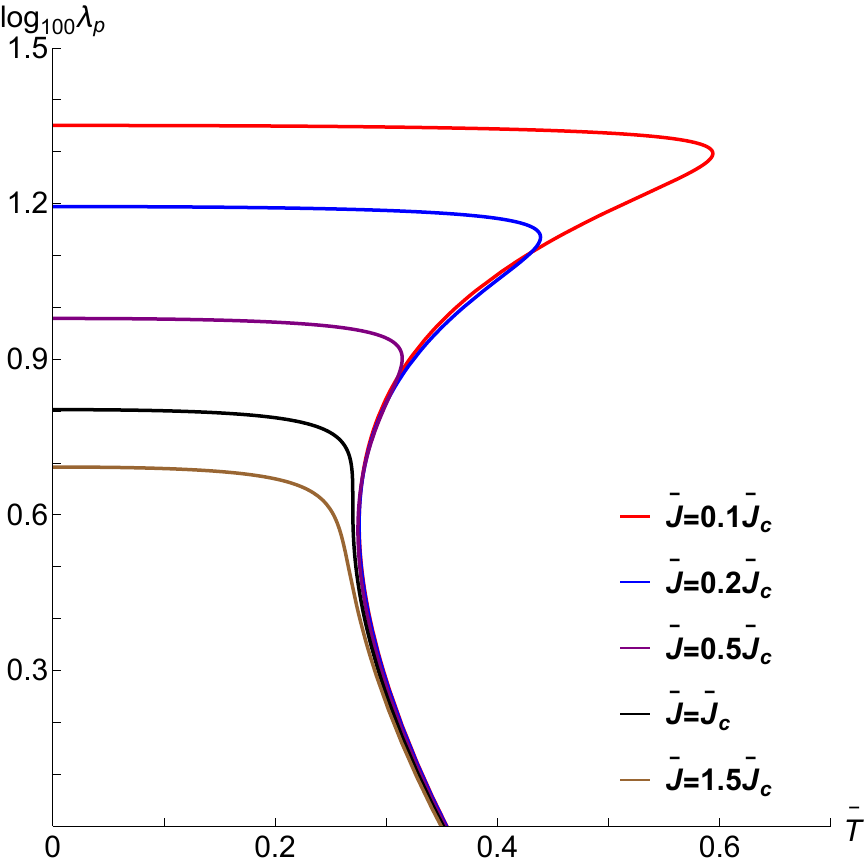}
			\subcaption{}
		\label{4-3-a}
	\end{minipage}
	\begin{minipage}[t]{0.48\textwidth}
		\centering
		\includegraphics[width=7cm,height=5cm]{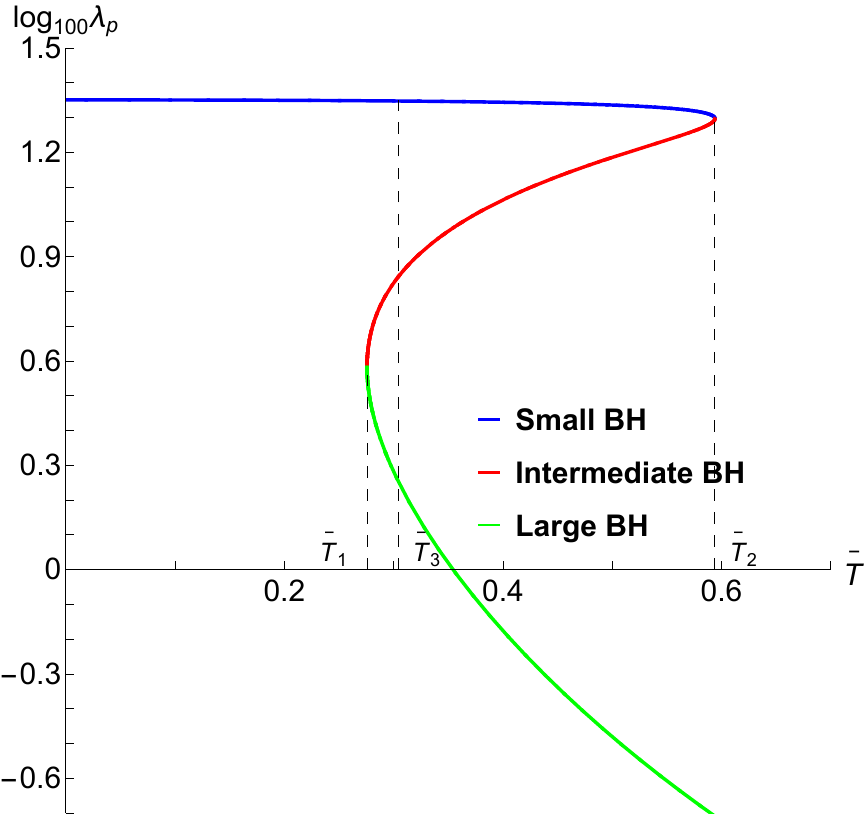}
			\subcaption{}
		\label{4-3-b}
	\end{minipage}
	\caption{The variation of the PTLE of chaos for the massless particle, as a function of the temperature of the four-dimensional Kerr-AdS BH. The specific scenario with $\bar{J}=0.1\bar{J}_c$ is depicted in \ref{4-3-b}.}
	\label{4f3}
\end{figure}

\begin{figure}[h]
	\centering
	\begin{minipage}[t]{0.48\textwidth}
		\centering
		\includegraphics[width=7cm,height=5cm]{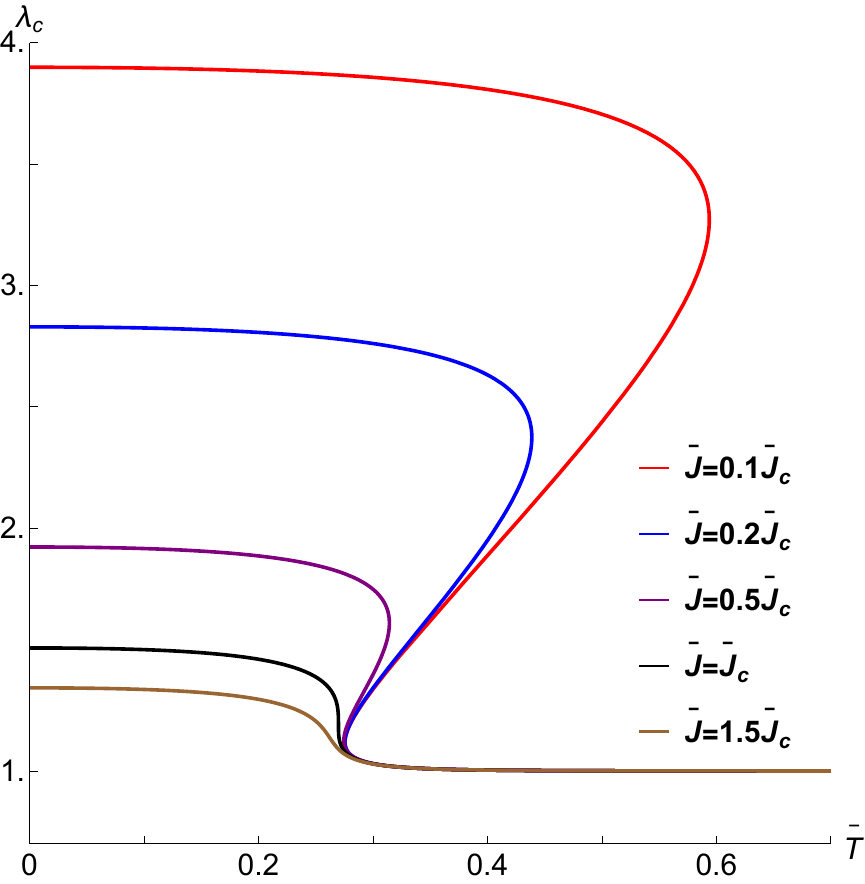}
		\subcaption{}
		\label{4-4-a}
	\end{minipage}
	\begin{minipage}[t]{0.48\textwidth}
		\centering
		\includegraphics[width=7cm,height=5cm]{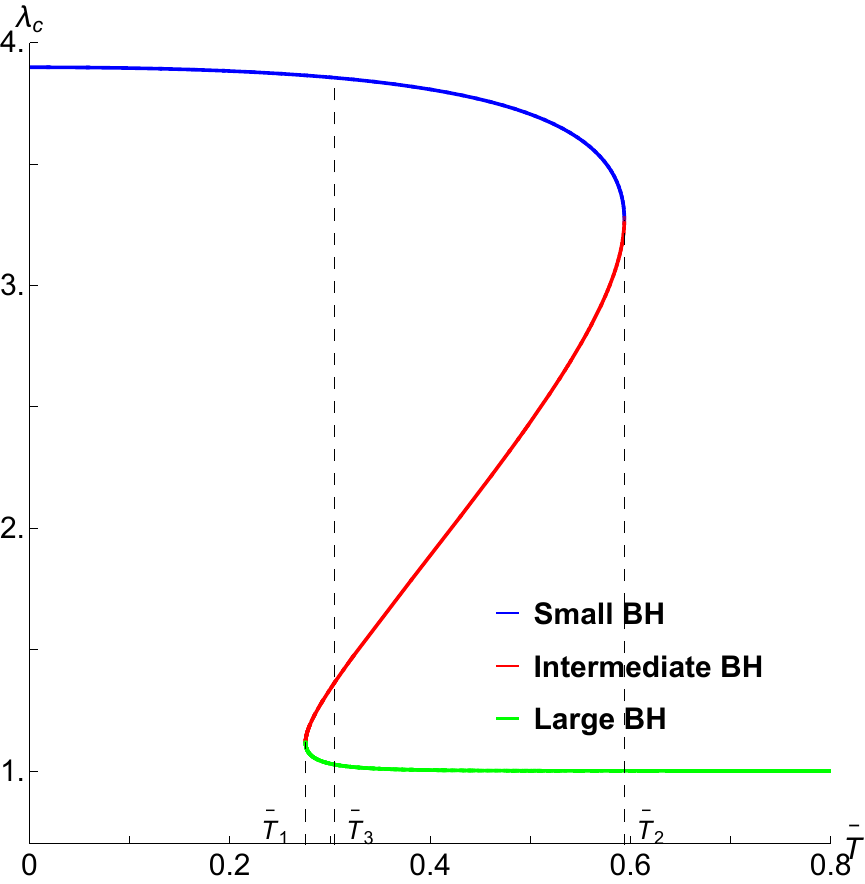}
		\subcaption{}
		\label{4-4-b}
	\end{minipage}
\caption{The variation of the CTLE of chaos for the massless particle, as a function of the temperature of the four-dimensional Kerr-AdS BH. The specific scenario with $\bar{J}=0.1\bar{J}_c$ is depicted in \ref{4-4-b}.}
	\label{4f4}
\end{figure}

\noindent We initially determine the position of the unstable equilibrium orbit for the particle, which is governed by the conditions $V_{r}=V_{r}^{\prime}=0$ and $V_{r}^{\prime\prime}>0$. Subsequently, we study the relationship between the PTLE and CTLE of chaos for the particle at this orbit and the temperature. Utilizing Eqs. (\ref{eq2.1.10}), (\ref{eq2.1.11}), (\ref{eq2.2.3}) and (\ref{eq2.3.1}), we obtain the PTLE and CTLE. Their variations with the temperature are illustrated in Figures \ref{4f3} and \ref{4f4}, respectively. For the calculations presented in this letter, we order $L=20l$.

The variation of the PTLE with the temperature is depicted in Figure \ref{4f3}. Specifically, in Figure \ref{4-3-a}, the exponent decreases monotonically when $\bar{J}=1.5\bar{J}_c$, whereas it is a multivalued function with respect to $\bar{T}$ when $\bar{J}<\bar{J}_c$ (where $\bar{J}_c$ is defined in Section \ref{sec3.1}). Figure \ref{4-3-b} illustrates the exponent's variation with the temperature for the case when $\bar{J}=0.1\bar{J}_c$. Here, it decreases monotonically when  $\bar{T} < \bar{T}_1$ and $\bar{T} >\bar{T}_2$, and assumes three values for a specific value of $\bar{T}$ within the range $\bar{T}_1 <\bar{T} <\bar{T}_2$ (where $\bar{T}_1$ and $\bar{T}_2$ correspond exactly to the temperatures in Figure \ref{4f2}, respectively). The small and large BH branches emerge in the ranges $\bar{T} < \bar{T}_1$ and $\bar{T} >\bar{T}_2$ respectively, and they are stable.  Between  $\bar{T}_1$ and $\bar{T}_2$, the small, intermediate and large BH phases coexist, implying that they can transform into each other within this temperature range. The first order phase transition occurs at $\bar{T}_3$. These behaviors resemble those observed in Figure \ref{4f2}, thereby indicating that the PTLE serves as a probe for the phase transition of the four-dimensional Kerr-AdS BH.

The curves in Figure \ref{4f4} illustrates the variation of the CTLE with the temperature for different values of  $\bar{J}$. In Figure \ref{4-4-a}, the exponent is a multivalued function when $\bar{J}<\bar{J}_c$, whereas it consistently assumes a single value for a specific temperature when $\bar{J}>\bar{J}_c$. As the temperature increases, the exponents for different values of $\bar{J}$ eventually converge, indicating a pronounced influence of the temperature on the exponent. The graph in Figure \ref{4-4-b} depicts the trend of the exponent as it varies with the temperature when $\bar{J}=0.1\bar{J}_c$. It is evidently that, for a given temperature, the exponent is multivalued within the range $\bar{T}_1 < \bar{T} < \bar{T}_2$ and single valued when $\bar{T} < \bar{T}_1$ and $\bar{T} > \bar{T}_2$. Within the range $\bar{T}_1 < \bar{T} < \bar{T}_2$, two scenarios emerge: in the small and large BH branches, the exponent decreases with increasing $\bar{T}$, whereas it increases with increasing $\bar{T}$ in the intermediate BH branch. Comparison with Figure \ref{4f2} reveals that the temperature region with multiple values of the exponent corresponds to the phase transition region depicted in Figure \ref{4f2}. In the large BH branch, where $\bar{T} > \bar{T}_2$, the exponent approaches $1$ as $\bar{T}$ continues to increase. Therefore, the CTLE also serves as a probe for the phase transition.

\subsubsection{Timelike geodesic's case}\label{sec3.2.2}

To explore the relationship between the phase transition and the PTLE of chaos for the massive particles, as well as its relationship with the CTLE, we undertake the calculation of the effective potential for such particles. Specifically, for a time-like geodesic, where $\delta =-1$, the corresponding effective potential is expressed as follows

\begin{eqnarray}
V_{r} = \frac{-\Delta r^2+(aE-L\Xi)^2\Delta-\left[E(r^2+a^2)-aL\Xi_a)\right]^2}{r^4}.
\label{eq2.3.2}
\end{eqnarray}

\begin{figure}[h]
	\centering
	\begin{minipage}[t]{0.48\textwidth}
		\centering
		\includegraphics[width=7cm,height=5cm]{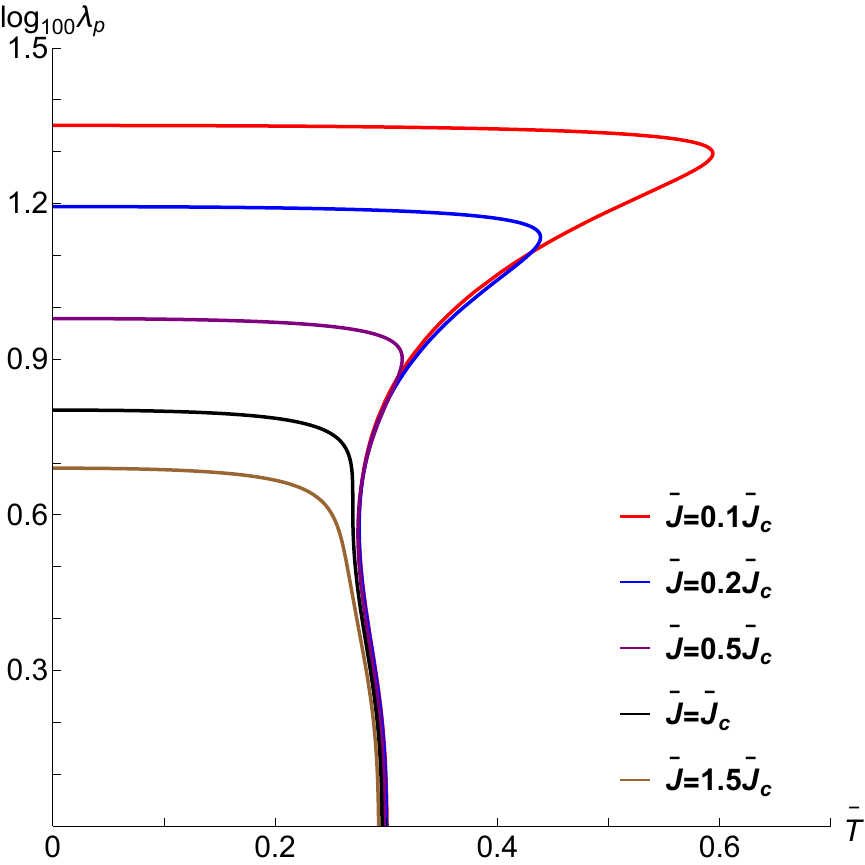}
		\subcaption{}
		\label{4-5-a}
	\end{minipage}
	\begin{minipage}[t]{0.48\textwidth}
		\centering
		\includegraphics[width=7cm,height=5cm]{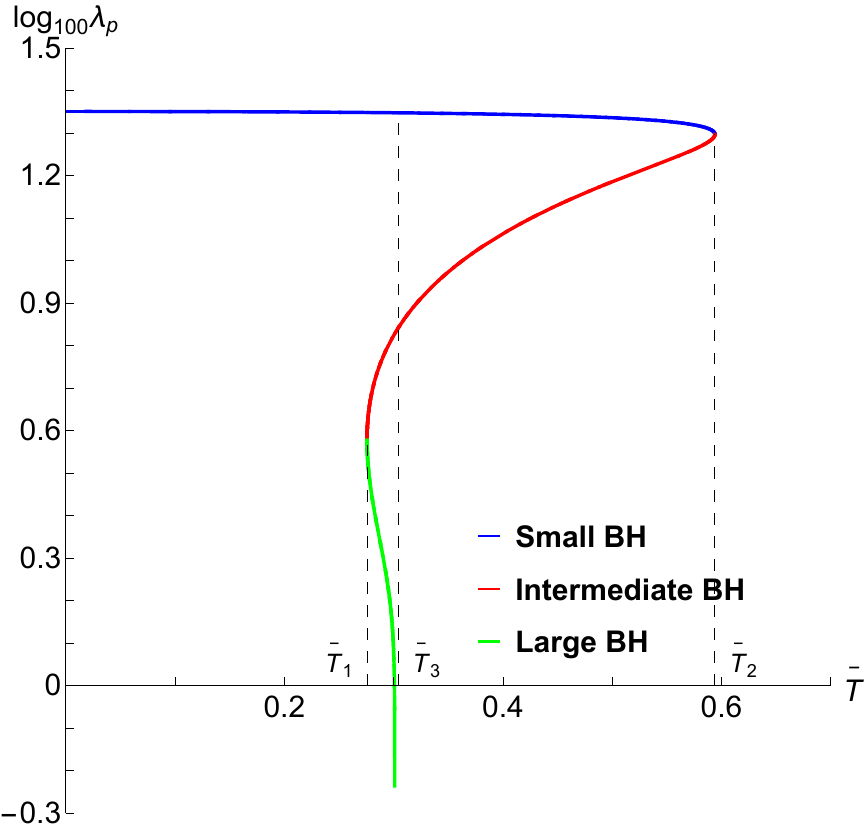}
		\subcaption{}
		\label{4-5-b}
	\end{minipage}
\caption{The variation of the PTLE of chaos for the massive particle, as a function of the temperature of the four-dimensional Kerr-AdS BH. The specific scenario with $\bar{J}=0.1\bar{J}_c$ is depicted in \ref{4-5-b}.}
	\label{4f5}
\end{figure}

\begin{figure}[h]
	\centering
	\begin{minipage}[t]{0.48\textwidth}
		\centering
		\includegraphics[width=7cm,height=5cm]{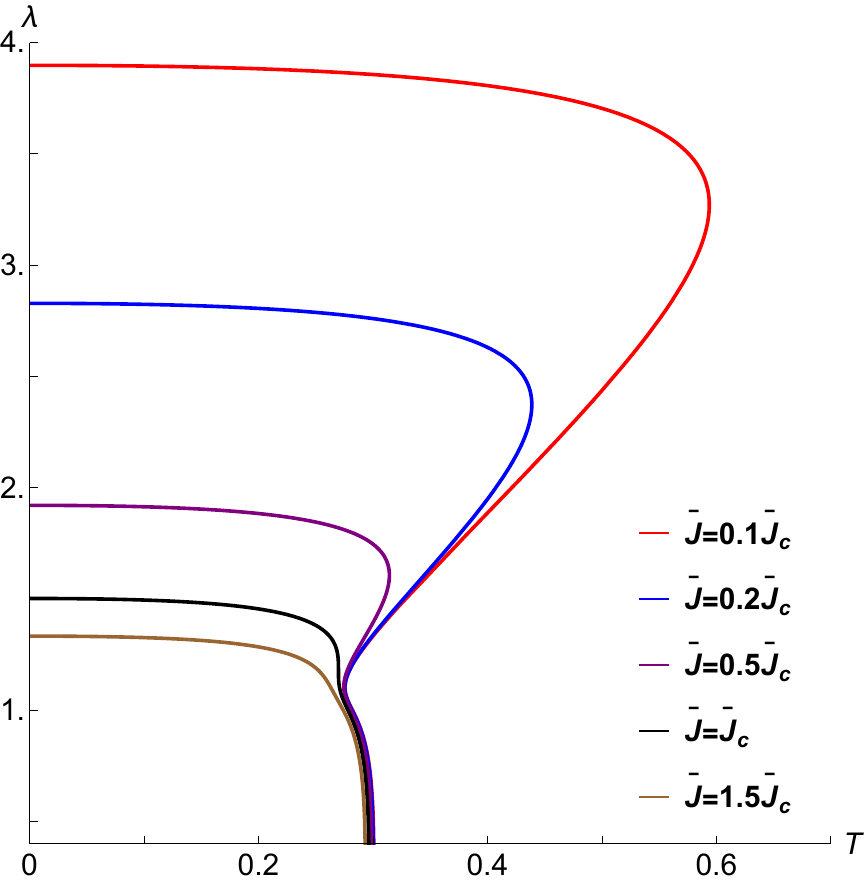}
		\subcaption{}
		\label{4-6-a}
	\end{minipage}
	\begin{minipage}[t]{0.48\textwidth}
		\centering
		\includegraphics[width=7cm,height=5cm]{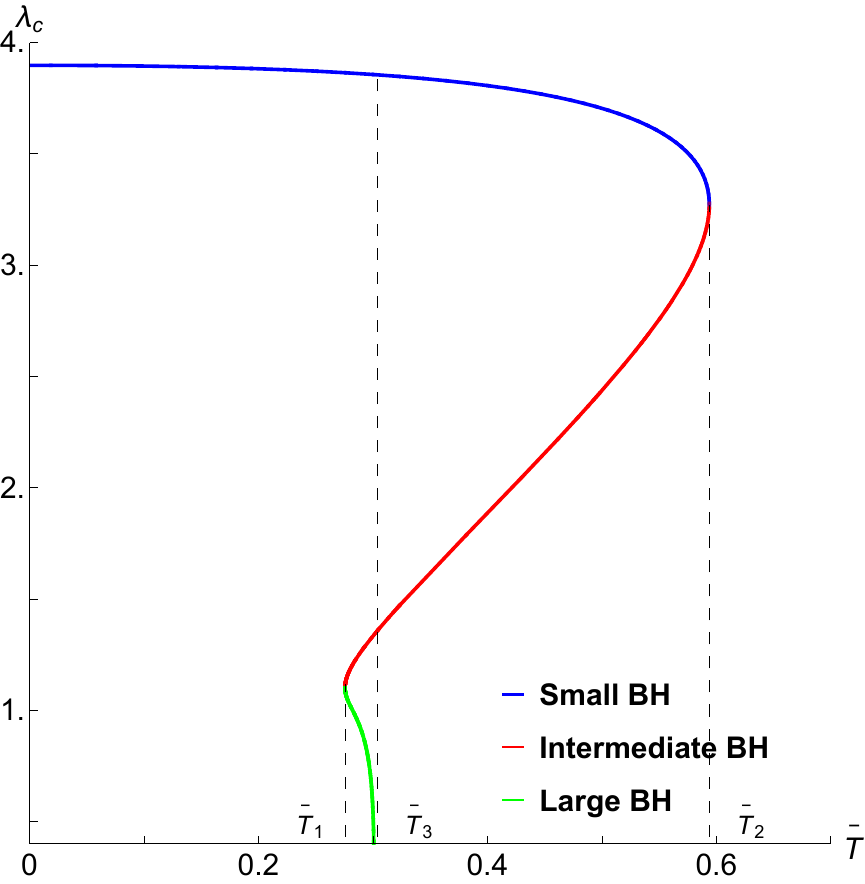}
		\subcaption{}
		\label{4-6-b}
	\end{minipage}
	\caption{The variation of the CTLE of chaos for the massive particle, as a function of the temperature of the four-dimensional Kerr-AdS BH. The specific scenario with $\bar{J}=0.1\bar{J}_c$ is depicted in \ref{4-6-b}.}
	\label{4f6}
\end{figure}

\noindent Thus the position of the unstable equilibrium orbit for the massive particle is determined by solving the equations  $V_{r}=V_{r}^{\prime}=0$ and $V_{r}^{\prime\prime}>0$. Subsequently, by utilizing Eqs. (\ref{eq2.1.10}), (\ref{eq2.1.11}), (\ref{eq2.2.3}) and (\ref{eq2.3.2}), we obtain the PTLE and CTLE at this orbit. The variations of these quantities with the temperature are depicted in Figures \ref{4f5} and \ref{4f6}, respectively.

Figure \ref{4f5} depicts the influence of the temperature on the PTLE for different values of $\bar{J}$. In Figure \ref{4-5-a}, we similarly observe that the exponent is a multivalued function when $\bar{J}<\bar{J}_c$, hinting at the presence of phase transitions. The scenario for $\bar{J}=0.1\bar{J}_c$ is illustrated in Figure \ref{4-5-b}, where $\bar{T}_{P1}$ signifies a terminal temperature for the exponent. At this temperature, the unstable equilibrium orbit disappears and the exponent is not equal to zero. When $\bar{T}<\bar{T}_1$, the exponent decreases monotonically with increasing the temperature. However, within the range  $\bar{T}_1 <T<\bar{T}_2$, the exponent becomes a multivalued function of the temperature. Specifically, within $\bar{T}_1< \bar{T}< \bar{T}_{p1}$, each temperature corresponds to three distinct values of the exponent, which represent the large, intermediate and small BH phases, respectively. This indicates that within this temperature range, these phases coexist, with the intermediate BH branch being unstable. Within $\bar{T}_{p1}< \bar{T}< \bar{T}_2$, each temperature maps to two values of the exponent, associated with the intermediate and small BH phases. Due to the emergence of the terminal temperature, the large BH branch cannot be described by the exponent within the temperature range from $\bar{T}_1$ to $\bar{T}_2$.

In Figure \ref{4f6}, the relationship between the CTLE of chaos for the massive particle and the temperature is plotted. Figure \ref{4-6-a} exhibits similarities to Figure \ref{4-5-a}. Specifically, when $\bar{J}<\bar{J}_c$, the exponent is a multivalued function of $\bar{T}$. Furthermore, there exists a point where the exponent equals zero for different values of $\bar{J}$, indicating the disappearance of the unstable equilibrium orbit. Figure \ref{4-6-b} presents the specific scenario for $\bar{J}=0.1\bar{J}_c$. When $\bar{T}=\bar{T}_{p2}$, the unstable equilibrium orbit disappears and the exponent is zero. Within the range $\bar{T}_1< \bar{T}< \bar{T}_{p2}$, the exponent remains a multivalued function of $\bar{T}$, with the large, intermediate and small BH phases coexisting within this temperature range. For a given temperature $\bar{T}$ within $\bar{T}_{p2}<\bar{T}<\bar{T}_2$, the exponent takes two values, which correspond to the intermediate and small BH phases, respectively. Therefore, the exponent of chaos for the massive particle also reveals the phase transition of the Kerr-AdS BH.

\subsubsection{Critical exponents}\label{sec3.2.3}

Upon examining critical points, it is often observed that certain physical quantities exhibit exponential growth or decay, with the rate of this change being characterized by critical exponents. This phenomenon is typically studied through heat capacity, where discontinuities in the heat capacity are identified as critical points. Various types of phase transitions possess unique critical exponents. Therefore, by analyzing these exponents, one can distinguish their differentiation and study their underlying  physical mechanisms. In the following, we employ  an elegant approach proposed in \cite{RBDR1,RBDR2} to calculate critical exponents which are related to the CTLE and PTLE. In their work, the heat capacity at constant charge is given by $C_Q= T \left(\frac{\partial S}{\partial T}\right)_Q = T \frac{\left(\partial S/\partial r_+\right)_Q}{\left(\partial T/\partial r_+\right)_Q}$, and the critical points are determined by the condition $\left(\frac{\partial T}{\partial r_+}\right)_Q=0$. For the case of the four-dimensional Kerr-AdS BH, the critical points are located at $\bar{r}_i$ and are determined by $\left(\frac{\partial \bar{T}}{\partial \bar{r}_+}\right)_{\bar{J}}=0$. 

Near a critical point, the horiozn radius is written as

\begin{eqnarray}
 \bar{r}_+=\bar{r}_i(1+\epsilon),
\label{eq2.4.1}
\end{eqnarray}

\noindent where $\bar{r}_i$ is the horizon radius at the critical point and $|\epsilon|<<1$. The rotational parameter can be written as a function about $\bar{r}_+$. Thus it is expressed as   

\begin{eqnarray}
    \bar{J}(\bar{r}_+)=\bar{J}_i(1+\zeta),
\label{eq2.4.2}
\end{eqnarray}

\noindent where $|\zeta|<<1$. We perform Taylor expansion on $\bar{J} (\bar{r}_+)$ within a sufficiently small neighborhood of $\bar{r}_i$ and  obtain

\begin{eqnarray}
    \bar{J}(\bar{r}_+)=\bar{J}(\bar{r}_i)+\left(\frac{\partial \bar{J}}{\partial \bar{r}_+}  \right)_c\left(\bar{r}_+-\bar{r}_i \right)+\frac{1}{2}\left(\frac{\partial^2 \bar{J}}{\partial \bar{r}^2_+} \right)_c\left(\bar{r}_+-\bar{r}_i \right)^2+ \mathcal{O}(\bar{r}_i).
\label{eq2.4.3}
\end{eqnarray}

\noindent In this subsection, the subscript ‘c’ represents values at the critical point. It is clearly that at this point, $\left(\frac{\partial \bar{J}}{\partial \bar{r}_+}  \right)_c =\left(\frac{\partial \bar{J}}{\partial \bar{T}} \right)_c \left(\frac{\partial \bar{T}}{\partial \bar{r}_+} \right)_c= 0$. Therefore, the second term on the right hand side of the above equation disappears. $ \mathcal{O}(\bar{r}_i)$ denotes all the higher order terms and is neglected. Using Eqs. (\ref{eq2.4.1}) and (\ref{eq2.4.3}), we get

\begin{eqnarray}
    \epsilon^2= \frac{1}{2} \frac{\bar{J}_i\zeta }{\bar{r}^2_i}\left(\frac{\partial^2 \bar{J}}{\partial \bar{r}^2_+} \right)_c.
\label{eq2.4.4}
\end{eqnarray}

\noindent We use $\lambda $ here to represent both the CTLE and PTLE. Performing Taylor expansion on $\lambda (\bar{r}_+)$ close to the critical point $\bar{r}_i$ yielding 

\begin{eqnarray}
    \lambda(\bar{r}_+)=\lambda(\bar{r}_i)+\left(\frac{\partial \lambda}{\partial \bar{r}_+}\right)_c\left(\bar{r}_+-\bar{r}_i\right)+ \mathcal{O}(\bar{r}_i).
\label{eq2.4.5}
\end{eqnarray}

\noindent  Ignoring all the higher order terms in the above equation and using Eqs. (\ref{eq2.4.2}), (\ref{eq2.4.4}) and (\ref{eq2.4.5}), we obtain

\begin{eqnarray}
    \lambda(\bar{r}_+)-\lambda(\bar{r}_i)=\left(\frac{\partial \lambda}{\partial \bar{r}_+}\right)_c\left(\frac{1}{2}\frac{\partial^2 \bar{J}}{\partial \bar{r}^2_+} \right)^{-\frac{1}{2}}_{\bar{r}_+=\bar{r}_i}\left(\bar{J}-\bar{J}_i\right)^{\frac{1}{2}}.
\label{eq2.4.6}
\end{eqnarray}

\noindent When a critical exponent $\bar{\delta}_1$ is defined as $\Delta\lambda \sim |\bar{J}-\bar{J}_c|^{\bar{\delta}_1}$, $\bar{\delta}_1$ is obtained as $1/2$. By using similar calculations as above, we obtain

\begin{eqnarray}
 \lambda(\bar{r}_+)-\lambda(\bar{r}_i)=\left(\frac{\partial \lambda}{\partial \bar{r}_+}\right)_c\left(\frac{1}{2}\frac{\partial^2 \bar{T}}{\partial \bar{r}^2_+} \right)^{-\frac{1}{2}}_{\bar{r}_+=\bar{r}_i}\left(\bar{T}-\bar{T}_i\right)^{\frac{1}{2}}.
\label{eq2.4.7}
\end{eqnarray}

\noindent From the definition of the critical exponent $\bar{\delta}_2 $ which satisfies $\Delta\lambda \sim |\bar{T}-\bar{T}_c|^{\bar{\delta}_2}$, $\bar{\delta}_2$ is  also derived as $1/2$. This result aligns with the findings in \cite{GLMW,LTW,SDDM}, where in their work, $\Delta\lambda =\lambda_s-\lambda_l $ is served as an order parameter, and $\lambda_s$ and $\lambda_l$ are the LEs of the small and large BHs, respectively.

\section{Conclusions}

In this letter, we studied the PTLEs and  CTLEs of chaos for both massless and massive particles orbiting the four-dimensional Kerr-AdS BH, and explored their relationships with the phase transition of this BH. When examining the relationships between these exponents and the temperature, we observed that the exponents are single valued functions of the temperature when the BH's angular momentum exceed its critical value. Conversely, when the angular momentum falls below the critical value, the exponents transform into the multivalued functions of the temperature. By comparing this behavior with the relationship between the free energy and the temperature, we found that these two types of LEs are capable of revealing the phase transition of the four-dimensional Kerr-AdS BH. Nevertheless, the PTLE and CTLE of chaos for the massless particle offer a more effective depiction of the phase transition compared to those of the massive particle. The critical exponents associated with the CTLEs and PTLEs were found to be 1/2.

\begin{acknowledgments}
We greatly appreciate the anonymous reviewer for the insightful comments that improved this work greatly. We would like to thank Dr. Chuanhong Gao and Dr. Xin Lv for their useful discussions.

\end{acknowledgments}

\end{document}